\documentstyle[psfig,aps]{revtex}
\begin{document}
\def\v{\rule{.4mm}{2.4mm}}
\def\h{\rule[1mm]{2.4mm}{.4mm}}
\draft
\title{Toward a Theory of Orbiton Dispersion in LaMnO$_3$}
\author{Vasili Perebeinos and Philip B. Allen}
\address{Department of Physics and Astronomy, State University of New York,
Stony Brook, New York 11794-3800}
\date{\today}
\maketitle
\begin{abstract}

At 750K, LaMnO$_3$ has a cooperative Jahn-Teller (JT) distortion,
with Mn atoms in distorted oxygen octahedra. This lifts the
degeneracy of the singly-occupied $e_g$ orbitals of the Mn$^{3+}$ ions,
which then become orbitally ordered.  We use a minimal model to describe
the ordered phase at $T=0$.  The on-site Coulomb repulsion $U$ is set to 
infinity.  There are two electronic orbitals and three oxygen
vibrational coordinates per unit cell.  In addition to spin excitations
and phonons, the model has electronic excitations consisting of
mis-orienting orbitals on Mn ions.  Neglecting coupling to the
oxygen displacements, the gap to such excitations is $2\Delta=16g^2/K$
where $g$ is the electron-phonon coupling and $K$ is the oxygen spring
constant.  When static oxygen displacements are coupled, this
excitation becomes a self-trapped exciton with energy $\Delta$, half
the JT gap.  Adding dynamic oxygen displacements in one-phonon approximation
introduces dispersion to both the (previously Einstein-like) phonons and
the orbital excitons (``orbitons'').  One of the phonon branches 
has zero frequency at $\vec{k}=(0,0,\pi)$.  This is the Goldstone
mode of the JT broken symmetry.

\end{abstract}

\section{introduction}

The fascinating properties of Manganite perovskites have recently been 
rediscovered \cite{Jonker,Kusters,Jin}. 
LaMnO$_3$ has Mn$^{3+}$ ions with electronic configuration
$3d^4$ in the high-spin $t_{2g\uparrow}^3 e_{g\uparrow}^1$ state.
The single electron in the doublet $e_g$ level is Jahn-Teller (JT)
active.  A cubic to orthorhombic distortion \cite{Goodenough} occurs
at $T_{\rm JT}$=750K which lifts the degeneracy with alternating sign in
alternating cells, and drives orbital ordering \cite{Murakami}: 
$x$ and $y$-oriented $e_g$ orbitals alternate in the $x-y$ plane.  This 
cooperative JT transition yields a layer structure
which then determines at the much lower Neel temperature ($T_{\rm N}$=240K)
an antiferromagnetic ``A-type'' (AFA) \cite{Wollan} 
spin-ordering with ferromagnetic
spin order in the $x-y$ planes.  We believe that the principal term
in the Hamiltonian which drives the cooperative JT transition is
coupling of Mn orbitals to motions of oxygen atoms along the directions
of the (approximately 180$^{\circ}$) Mn-O-Mn bonds.  In a previous
paper \cite{AP} we have discussed a model for the
JT transition and used this model both to describe the $T=0$
phase of pure LaMnO$_3$ and the formation of small polarons when
hole-doped as in La$_{1-x}$Ca$_x$MnO$_3$.  Our model is
related to the Rice-Sneddon \cite{Rice} model for BaBiO$_3$, 
and was introduced for LaMnO$_3$ by Millis \cite{Millis}. 

\begin{figure}
\centerline{
\psfig{figure=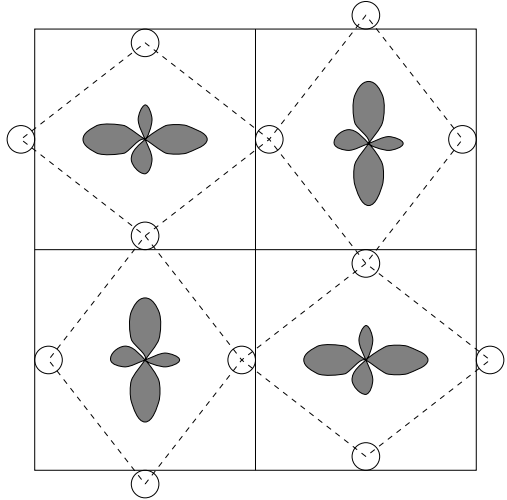,height=2.6in,width=2.6in,angle=0}}
\caption{Schematic of Jahn-Teller distorted LaMnO$_3$ $x-y$ plane
with orbitals ordered}
\label{JT}
\end{figure}

Our calculations are done at $T$=0 and $x$=0.  
Previous authors \cite{Ishihara,Feiner,Horsch} have studied orbital
dynamics using models which omit JT and electron-phonon effects, 
focussing on dynamics induced by hopping.  This leads to complicated
orbital dynamics strongly dependent on spin order because of Hubbard
and Hund energies.  Our model emphasizes a very different kind
of physics.  Because we take $U=\infty$ and $x$=0,
there is no hopping and the spins are decoupled from other degrees of
freedom.  The remaining electron-phonon physics is still very rich, and
we think it will be dominant at small $x$.  Other treatments of
the small $x$ orbital dynamics exist
(see ref. \cite{Brink} and papers cited there), but the large
ratio $T_{\rm JT}/T_{\rm N}$ argues against omitting the
electron-phonon physics at small $x$.

\begin{figure}
\centerline{
\psfig{figure=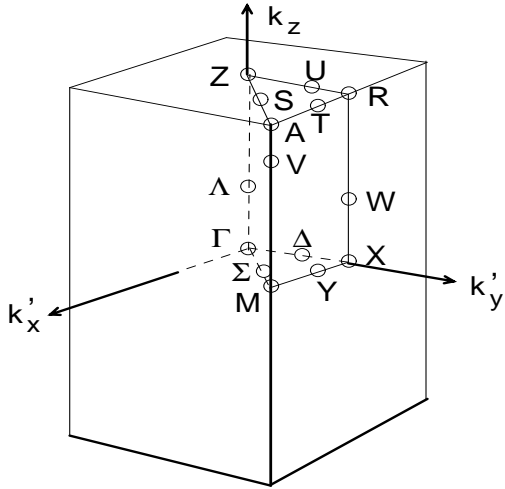,height=2.6in,width=2.6in,angle=0}}
\caption{Tetragonal Brillouin zone of the Jahn-Teller distorted 
phase of LaMnO$_3$} 
\label{BZ}
\end{figure}

\section{model and ground state}

Each cubic cell has 5 degrees of freedom, two electronic
and 3 lattice dynamical.  The electronic degrees of freedom are
occupancies of the Mn $e_g$ orbitals.   In Van Vleck notation \cite{Vleck}
$\psi_2$ and $\psi_3$ are the familiar $x^2-y^2$ and $3z^2-r^2$ functions.
These can be rotated by 45$^{\circ}$ to an alternative orthogonal basis, the
symmetrical pair $\psi_X$, $\psi_Y=(-\psi_3\pm\psi_2)/\sqrt{2}$.
These orbitals point in $x$ and $y$ directions respectively,
with a moderately strong lobe in the $z$ direction.  

Lattice dynamical degrees of freedom are labeled by
$u(\ell\pm x)$, $u(\ell\pm y)$, and $u(\ell\pm z)$, which refer to
oxygen displacements along the $x$, $y$, and $z$ directions of the
oxygen atoms half-way between the Mn atom at site $\ell$ and the
next atom in the $\pm x$, $\pm y$, and $\pm z$ directions, respectively.
The notation $u(\ell -x)$ is equivalent to $u(\ell^{\prime} +x)$ for
the Mn site $\ell^{\prime}$, which is translated from $\ell$ by one unit in the $-x$-direction.

It is convenient to use both the Cartesian labeling given above, and
a Van Vleck labeling:
\begin{eqnarray}
\sqrt{3/2} Q_1(\ell)&=&[u(\ell+x)- u(\ell-x)]+[u(\ell+y)- u(\ell-y)]
     +[u(\ell+z)-u(\ell-z)]\nonumber\\
Q_2(\ell)&=&[u(\ell+x)- u(\ell-x)] -[u(\ell+y)- u(\ell-y)]\nonumber\\
\sqrt{3} Q_3(\ell)&=&2[u(\ell+z)-u(\ell-z)]
   -[u(\ell+x)- u(\ell-x)] -[u(\ell+y)- u(\ell-y)]
\label{vvu}
\end{eqnarray}
Our starting Hamiltonian ${\cal H}$ has a lattice term ${\cal H}_{\rm L}$
and an electron-phonon term ${\cal H}_{\rm ep}$.
The lattice term does not have a useful expression in the Van Vleck
notation, and is best written as
\begin{equation}
{\cal H}_{\rm L}=\sum_{\ell, \lambda=x,y,z}\left(\frac{K}{2}u^2(\ell+\lambda)+
\frac{1}{2M}p^2(\ell+\lambda)\right).
\label{hlat}
\end{equation}
On the other hand, the electron-phonon term is best written in Van Vleck
operators,
\begin{eqnarray}
{\cal H}_{\rm ep}&=&-g\sum_{\ell}\left(c_X^{\dagger}(\ell),
c_Y^{\dagger}(\ell)\right)
\left( \begin{array}{rr}  Q_2(\ell) & -Q_3(\ell) \\
             -Q_3(\ell) & -Q_2(\ell)   \end{array} \right)
\left( \begin{array}{c} c_X(\ell)\\c_Y(\ell)
\end{array}\right)
\nonumber\\
&&-\sqrt{2}g\sum_{\ell}Q_1(\ell)\left(c_X^{\dagger}(\ell)c_X(\ell)+
c_Y^{\dagger}(\ell)c_Y(\ell)\right).
\label{hep}
\end{eqnarray}
In the undoped case, each Mn site is singly occupied; we stay in
the subspace where the operator relation
\begin{equation}
c_X^{\dagger}(\ell)c_X(\ell)+ c_Y^{\dagger}(\ell)c_Y(\ell)=1
\label{unitocc}
\end{equation}
holds.  The degenerate $e_g$ orbitals are defined to have zero energy.

In adiabatic approximation, the ground state of this Hamiltonian has 
a doubled unit cell and
a net average oxygen displacement $(Q_2(\ell),Q_3(\ell))$ 
of the type $Q(\cos\theta,\sin\theta)\exp(i\vec{q}\cdot\vec{\ell})$.
The wave-vector $\vec{q}=(\pi,\pi,\pi)$ is favored, but the angle
$\theta$ in the $(Q_2,Q_3)$-plane is undetermined.  The observed
distortion, of the $Q_2$-type ($\theta=0$), is fixed by terms left
out of our Hamiltonian.  Once a $Q_2$-type order parameter is
established, the wave-vector $\vec{q}$ can equally well be $(\pi,\pi,0)$,
and this is what is observed.  The lattice is now bi-partite.  We
label sites by A when $\exp(i\vec{q}\cdot\vec{\ell})$ is 1, and
by B when it is -1.  According to Eqn. \ref{hep}, the A sites
will be occupied by X orbitals and the B sites by Y orbitals.
The alternating occupancy of these orbitals on A and B sublattices
is shown schematically in Fig. (\ref{JT}).  

Oxygen atoms move off-center by $\pm 2g/K$, giving a JT distortion
$Q_2=8g/K\exp(i\vec{q}\cdot\vec{\ell})$.  This
lowers the energy of each occupied orbital by $8g^2/K=\Delta$ and 
raises the empty orbitals by the same amount.  The distortion
raises the lattice energy by $4Ng^2/K$, which is half the energy
lowering of $8Ng^2/K$ from Jahn-Teller energy.
The lowest electronic
excited states (``orbitons'') are misoriented orbitals which cost
$16g^2/K=2\Delta\approx$2eV.

\section{orbiton Hamiltonian}

We are interested in the lowest excited states of the model.  We allow 
vibrations of oxygen atoms around their distorted positions.  
Therefore we are no longer making the adiabatic approximation.  Lattice
vibrations will renormalize both the ground state and the orbiton excitations.
By flipping an orbital we create an orbital defect.  We can define a
creation operator $a^{\dagger}(\ell)$ for the orbital defect on the A site as
$c_Y^{\dagger}(\ell)c_X(\ell)$, and similarly on the B site
$b^{\dagger}(\ell)= c_X^{\dagger}(\ell)c_Y(\ell)$.  
One can show that these operators obey 
Fermi commutation relations in the subspace of singly occupied sites.  
Furthermore one can show that on the A 
sublattice number the operator of the misoriented orbitals $c_Y^{\dagger}(\ell)
c_Y(\ell)$ is the same as $a^{\dagger}(\ell)a(\ell)$; similarly on the B
sublattice $c_X^{\dagger}(\ell)c_X(\ell)$ is the same as
$b^{\dagger}(\ell)b(\ell)$.

Using these definitions we can rewrite the Hamiltonian Eqns. \ref{hlat},
\ref{hep} in terms of the lattice fluctuations
$u(\ell+\lambda)$ defined as $u_0(\ell+\lambda)+{\delta}u(\ell+\lambda)$,
where  $u_0(\ell+\lambda)=\pm2g/K$ in $x$ or $y$ direction and $0$ in the $z$ 
direction.
\begin{eqnarray}
4Ng^2/K+{\cal H}_{\rm ep}+{\cal H}_{\rm L}
= &2& \Delta\left(\sum_{\ell\in A}a^{\dagger}
(\ell)a(\ell)+\sum_{\ell\in B}b^{\dagger}(\ell)b(\ell)\right)
\nonumber\\
&+& \sum_{\ell,\lambda=x,y,z}\left(\frac{K}{2}\delta u^2(\ell+\lambda)+
\frac{1}{2M}p^2(\ell+\lambda)\right)
\nonumber\\
&+& 2g\left(\sum_{\ell\in A}\delta Q_2(\ell)a^{\dagger}
(\ell)a(\ell)-\sum_{\ell\in B}\delta Q_2(\ell)b^{\dagger}(\ell)b(\ell)\right)
\nonumber\\
&+& g\left(\sum_{\ell\in A}\delta Q_3(\ell)\left(a^{\dagger}(\ell)+a(\ell)
\right)+
\sum_{\ell\in B}\delta Q_3(\ell)\left(b^{\dagger}(\ell)+b(\ell)\right)\right)
\label{hrsp}
\end{eqnarray}

The next step is into Fourier space.  Define Fourier variables by
\begin{eqnarray}
a(\ell)&=&
\sqrt{\frac{2}{N}}\sum_k a_k e^{i\vec{k} \cdot \vec{\ell}}\nonumber \\
b(\ell)&=&
\sqrt{\frac{2}{N}}\sum_k b_k e^{i\vec{k} \cdot \vec{\ell}}\nonumber \\
\delta u(\ell+\lambda)&=&\sqrt{\frac{2}{N}}\sum_k \sqrt{\frac{\hbar}
{2M\omega}}\delta u_{k_{\lambda}}e^{i\vec{k} \cdot \vec{\ell}}
\label{fourier}
\end{eqnarray}
where $\delta u_{k_{\lambda}}=f_{k_{\lambda}}+f_{-k_{\lambda}}^ {\dagger}$, 
if $\ell$ is on the A sublattice, and otherwise
$\delta u_{k_{\lambda}}=g_{k_{\lambda}}+g_{-k_{\lambda}}^ {\dagger}$.
The Fourier-transformed Hamiltonian has the form
\begin{equation}
{\cal H}={\cal H}_{\rm orb}^0+{\cal H}_{\rm ph}^0+{\cal H}_{\rm 3}(\delta Q_3)
+{\cal H}_{\rm 2}(\delta Q_2).
\label{hfourier}
\end{equation}
The first two terms are unperturbed dispersionless orbitons and phonons 
with energy $2\Delta$ and $\hbar \omega$. 
\begin{eqnarray}
{\cal H}_{\rm orb}^0&=&2\Delta\sum_{k}\left(a^{\dagger}_ka_k+b^{\dagger}_kb_k
\right)
\nonumber\\
{\cal H}_{\rm ph}^0&=&\hbar \omega\sum_{k,\lambda=x,y,z}\left(f^{\dagger}_
{k_{\lambda}}f_{k_{\lambda}}+g^{\dagger}_{k_{\lambda}}g_{k_{\lambda}}\right)
\label{h0}
\end{eqnarray}
The last two terms give the coupling of orbiton density to quantum
fluctuations of the lattice of the
$\delta Q_3$-type and $\delta Q_2$-type.
\begin{eqnarray}
{\cal H}_{\rm 3}(\delta Q_3)&=&\kappa_3\sum_{k}
\left[\left(a^{\dagger}_k+a_{-k}
\right)\left(A_1(k)+A_1^{\dagger}(-k)\right)
+\left(b^{\dagger}_k+b_{-k}
\right)\left(A_2(k)+A_2^{\dagger}(-k)\right)\right]
\nonumber\\
{\cal H}_{\rm 2}(\delta Q_2)&=&\kappa_2\sum_{k}
\left[\left(B_1(k)+B_1^{\dagger}(-k)\right)\rho_a(k)
+\left(B_2(k)+B_2^{\dagger}(-k)\right)\rho_b(k)\right]
\label{hpert}
\end{eqnarray}
We are now explicitly using the $(\pi\pi 0)$ wavevector for Jahn-Teller 
oxygen distortions.  New operators $A$, $B$, are introduced to represent
the Fourier transformed $\delta Q_3(k)$ and $\delta Q_2(k)$ 
variables, and new operators $\rho_a$ and $\rho_b$ are used for
orbiton densities on A and B sites:
\begin{eqnarray}
A_1(k)&=&2f_{k_z}\left(1-e^{-ik_z}\right)-f_{k_x}-f_{k_y}
+g_{k_x}e^{-ik_x}+g_{k_y}e^{-ik_y}
\nonumber\\
A_2(k)&=&2g_{k_z}\left(1-e^{-ik_z}\right)-g_{k_x}-g_{k_y}
+f_{k_x}e^{-ik_x}+f_{k_y}e^{-ik_y}
\nonumber\\
B_1(k)&=&f_{k_x}-f_{k_y}-g_{k_x}e^{-ik_x}+g_{k_y}e^{-ik_y}
\nonumber\\
B_2(k)&=&g_{k_x}-g_{k_y}-f_{k_x}e^{-ik_x}+f_{k_y}e^{-ik_y} \nonumber \\
\rho_a(k)&=&\sqrt{\frac{2}{N}}\sum_q a_{k+q}^{\dagger}a_q \nonumber \\
\rho_b(k)&=&\sqrt{\frac{2}{N}}\sum_q b_{k+q}^{\dagger}b_q 
\label{ABdef}
\end{eqnarray}
The term due to $\delta Q_3$ couples orbitons to the phonon subspace with 
coupling constant $\kappa_3=g\left(\hbar/6M\omega\right)^{1/2}=$ 
$\left(\hbar\omega\Delta/48\right)^{1/2}$. 
Since the phonon energy is much smaller than the orbiton energy 
$(\hbar\omega\ll \Delta)$ this term can be treated perturbatively.
The term due to $Q_2$ couples phonon vibrations to 
orbiton densities with the coupling constant  
$\kappa_2=2g(\hbar/2M\omega)^{1/2}=(\hbar\omega\Delta/4)^{1/2}$. 
This term describes oxygen vibrations around new positions which 
relax by an amount proportional to the orbiton density $\rho$ in order to 
lower the energy of the orbiton excitation. 
To deal with term we make a displaced oscillator transformation. 
In place of the phonon operators $f_{k_{\lambda}}, g_{k_{\lambda}}$ defined 
above,
we will use the following operators:
\begin{eqnarray}
F_{k_x}&=&f_{k_x}+\frac{\kappa_2}{\hbar\omega}\left(\rho_a(-k)+\rho_b(-k)
e^{ik_x}\right)
\nonumber\\
F_{k_y}&=&f_{k_y}-\frac{\kappa_2}{\hbar\omega}\left(\rho_a(-k)+\rho_b(-k)
e^{ik_y}\right)
\nonumber\\
G_{k_x}&=&g_{k_x}-\frac{\kappa_2}{\hbar\omega}\left(\rho_a(-k)e^{ik_x}+
\rho_b(-k)\right)
\nonumber\\
G_{k_y}&=&g_{k_y}+\frac{\kappa_2}{\hbar\omega}\left(\rho_a(-k)e^{ik_y}+
\rho_b(-k)\right)
\nonumber\\
F_{k_z}&=&f_{k_z}\nonumber\\
G_{k_z}&=&g_{k_z}
\label{trans}
\end{eqnarray}
These new operators are bosonic.  They were chosen to complete squares in 
the ${\cal H}_{\rm ph}^0+{\cal H}_{\rm 2}(\delta Q_2)$ part of the 
Hamiltonian, which becomes
\begin{eqnarray}
{\cal H}_{\rm ph}^0&+&{\cal H}_{\rm 2}(\delta Q_2)=
\hbar \omega\sum_{k,\lambda= x,y,z}\left(F^{\dagger}_{k_{\lambda}}
F_{k_{\lambda}}
+G^{\dagger}_{k_{\lambda}} G_{k_{\lambda}}\right) \nonumber\\
&-&\frac{(\kappa_2)^2}{\hbar\omega}\sum_{k,\lambda=x,y}\left[
\left(\rho_a(k)+ \rho_b(k)e^{-ik_{\lambda}} \right)\times\left({\rm HC}\right)
+\left(\rho_a(k)e^{-ik_{\lambda}}+\rho_b(k)\right)\times\left({\rm HC}\right)
\right]
\label{renorm}
\end{eqnarray}
where HC stands for Hermitean conjugate.

We now truncate the Hilbert space to the subspace with
only one orbiton present.  Then the product of two adjacent
orbiton annihilation operators gives zero. 
Then Eqn. \ref{renorm} can be simplified further by noting that 
$\sum_{k}[\rho_a(k)\rho_a(-k)+\rho_b(k)\rho_b(-k)]$ is the same as
$\sum_{k}[a^{\dagger}(k)a(k)+b^{\dagger}(k)b(k)]$; cross terms are zero 
in our subspace.  The second term of Eqn. \ref{renorm} then 
combines with ${\cal H}_{\rm orb}^{\rm 0}$ Eqn. (\ref{h0}) to
renormalize the orbiton energy by
$-4(\kappa_2)^2/\hbar\omega=-\Delta$.  The orbiton energy is now
exactly half of the bare energy $2\Delta$.  The reason is obvious
in a site representation.  Flipping an orbital creates a local region
where an orbital is surrounded in the $xy$ plane by other orbitals
of the same orientation.  Oxygen atoms adjacent to the central
(flipped) orbital will now prefer to locate in undistorted
positions.  This means that rather than raising the local energy 
from $-\Delta$ to $+\Delta$ with the flip, we have instead raised
it from $-\Delta$ to 0.  This is the sole influence of $\delta Q_2$-type
fluctuations in the one-orbiton subspace.

Now we transform the rest of the Hamiltonian to the displaced
oscillator operators.  Substituting new 
phonon operators (\ref{trans}) into ${\cal H}_{\rm 3}(\delta Q_3)$ generates 
the same term as in (\ref{hpert}), but with old phonon operators replaced
with new ones, plus additional terms coupling orbiton operators to orbiton 
density operators.  The new Hamiltonian has the following form:
\begin{equation}
{\cal H}={\cal H}_{\rm orb}+{\cal H}_{\rm ph}+{\cal H}_{\rm 1}+
{\cal H}_{\rm 2}+{\cal H}_{\rm 3}
\label{hnew}
\end{equation}
where the first two terms are the new bare orbitons and phonons
\begin{eqnarray}
{\cal H}_{\rm orb}&=&\Delta\sum_{k}
\left(a_k^{\dagger}a_k+b_k^{\dagger}b_k\right)
\nonumber\\
{\cal H}_{\rm ph}&=&\hbar \omega\sum_{k,\lambda=
x,y,z}\left(F^{\dagger}_{k_{\lambda}}F_{k_{\lambda}}+G^{\dagger}_{k_{\lambda}}
G_{k_{\lambda}}\right).
\label{hbare}
\end{eqnarray}
There are now three coupling terms left,
\begin{eqnarray}
{\cal H}_{\rm 1}&=&\kappa_3\sum_{k}\left(a^{\dagger}_kA_1(k)+b^{\dagger}_k
A_2(k)+{\rm HC}\right)
\nonumber\\
{\cal H}_{\rm 2}&=&\kappa_3\sum_{k}\left(a^{\dagger}_kA_1^{\dagger}(-k)+
b^{\dagger}_kA_2^{\dagger}(-k)+{\rm HC}\right)
\nonumber\\
{\cal H}_{\rm 3}&=&\kappa_2^{0}\sum_{k}\left(\cos k_x-\cos k_y\right)
\left(a^{\dagger}_k\rho_b(-k)+b^{\dagger}_k\rho_a(-k)+{\rm HC}\right).
\label{final}
\end{eqnarray}
Here the $A$ and $B$ operators are the same as defined in (\ref{ABdef})  
after substition of the new phonon operators $F$ and $G$ in place
of $f$ and $g$.  The
new coupling constant $\kappa_2^0$ is $4\kappa_1\kappa_2/\hbar\omega=\Delta/
\sqrt{12}$.  The perturbation 
part of the Hamiltonian ${\cal H}_{\rm 1}+{\cal H}_{\rm 2}+{\cal H}_{\rm 3}$ 
couples orbitons and phonons to states with energies lying above or below 
by the order of $\Delta$. These terms will be handled in the next 
section.

\section{Low energy Effective Hamiltonian}

In the Hamiltonian (\ref{hbare},\ref{final}) orbitons and phonons are coupled 
directly through the ${\cal H}_{\rm 1}$ term. 
The term ${\cal H}_{\rm 3}$ does not couple to the phonons, 
but couples the one orbiton subspace to the two orbiton
subspace.  The remaining ${\cal H}_{\rm 2}$ piece couples 
the one orbiton and one phonon states to states with one more orbiton 
and one more phonon.  Thus the Hamiltonian matrix acting on
the one phonon and one orbiton subspace has the structure: 

\begin{equation}
{\cal H}\Psi=\left( \begin{array}{ccccc}  
           \hbar \omega \hat I & {\cal H}_1 & {\cal H}_{2a}  & 0 & 0  \\
           {\cal H}_1 & \Delta \hat I & 0 & {\cal H}_3  & {\cal H}_{2b}   \\
           {\cal H}_{2a} & 0 & \left(\Delta+2\hbar \omega \right) \hat I & 0 
             & 0   \\
           0 & {\cal H}_3  & 0 &  2\Delta \hat I & 0  \\
           0 & {\cal H}_{2b} & 0 & 0 & 
               \left(2\Delta+\hbar \omega \right) \hat I  \\
              \end{array} \right)
                \left( \begin{array}{l}
			\Psi({\rm 1 ph})\\
			\Psi({\rm 1 orb})\\
			\Psi({\rm 1 orb+2 ph})\\
			\Psi({\rm 2 orb})\\
			\Psi({\rm 2 orb}+1ph)
                \end{array} \right)
\label{shcem}
\end{equation}
The operator ${\cal H}_{2}$ has matrix elements ${\cal H}_{2a}$
when it couples the 1 phonon to the 1 orbiton plus 2 phonon subspace,
and matrix elements ${\cal H}_{2b}$ when it couples the 1 orbiton to the
2 orbiton plus 1 phonon subspace.  The operator ${\cal H}_1$
couples 1 orbiton to 1 phonon.  It also couples 1 orbiton plus 2 phonons
to 2 orbitons plus 1 phonon.  However, this block does not influence
the 1 phonon or 1 orbiton answers except in higher order, so these
elements of the Hamiltonian matrix are omitted.

This Hamiltonian can be perturbatively reduced 
to two separate low-energy effective Hamiltonians, one for single orbitons 
and one for single phonons:

\begin{eqnarray}
{\cal H}_{\rm orb}^{\rm eff}&=&{\cal H}_{\rm orb}+{\cal H}_{\rm 1}\left(
E-\hbar \omega \right)^{-1}{\cal H}_{\rm 1}+
{\cal H}_{\rm 2b}\left(E-\left(2\Delta+\hbar \omega \right) \right)^{-1}
{\cal H}_{\rm 2b}
+{\cal H}_{\rm 3}\left(E-2\Delta \right)^{-1}{\cal H}_{\rm 3}
\nonumber\\
{\cal H}_{\rm ph}^{\rm eff}&=&{\cal H}_{\rm ph}+{\cal H}_{\rm 1}\left(
E-\Delta \right)^{-1}{\cal H}_{\rm 1}+
{\cal H}_{\rm 2a}\left(
E-\left(\Delta+2\hbar \omega \right) \right)^{-1}{\cal H}_{\rm 2a}
\label{ef}
\end{eqnarray}

The $2\times 2$ effective Hamiltonian ${\cal H}_{\rm orb}^{\rm eff}$
acts on wavefunctions built from two 
types of orbitons $\left(a_q^{\dagger},b_q^{\dagger}\right)\mid 0,0>$. 
Similarly the $6\times 6$ effective Hamiltonian 
${\cal H}_{\rm ph}^{\rm eff} $ acts on six-component 
wavefunctions $\left(F_{q_{\lambda}}^{\dagger},
G_{q_{\lambda}}^{\dagger}\right)\mid 0,0>$,where $\lambda=x, y, z$. 
The unperturbed ground state 
$\mid 0,0>$ with zero orbitons and phonons couples through the ${\cal H}_2$ 
term to states of one phonon and one orbiton.  This leads to a ground 
energy shift $\delta E_{GS}=<0,0 \mid {\cal H}_2 \left(E-\left(\Delta+
\hbar \omega \right)\right)^{-1}{\cal H}_2\mid 0, 0 >$. 

The problem is now reduced to computation and diagonalization of a $2\times 2$
and a $6\times 6$ matrix.
Note that we could have taken into account the ${\cal H}_1$ part 
of the perturbation Hamiltonian by leaving it as an off-diagonal coupling
term, giving an $8\times 8$ effective Hamiltonian to compute and
diagonalize.  The error we make by folding it down as in (\ref{ef})
has the same order of magnitude 
$(\hbar \omega / \Delta)^2$ as the other errors we make in our perturbation
theory. 

Our next task is to evaluate the matrix elements
of the effective Hamiltonians (\ref{ef}). After 
straight-forward algebra the result (with the ground energy shift 
$\delta E_{GS}=-12\kappa_3^{2}N/\left(\Delta+\hbar \omega \right)$ 
subtracted out) is:
\begin{eqnarray}
{\cal H}_{\rm orb}^{\rm eff}&=&\left(\Delta-\frac{(\kappa_2^0)^2}{\Delta}
\right)\hat I+
\left(\frac{\kappa_3^2}{\Delta-\hbar \omega}+\frac{\kappa_3^2}{\Delta+\hbar 
\omega}\right)
\left(\left(12-8\cos k_z\right)\hat I-2\left(\cos k_x+\cos k_y\right)\hat{\sigma}_x 
\right)
\nonumber\\
{\cal H}_{\rm ph}^{\rm eff}&=&\hbar \omega \hat I+\left(\frac{\kappa_3^2}
{\Delta-\hbar \omega}+\frac{\kappa_3^2}{\Delta+\hbar\omega}\right)
\left( \begin{array}{cc}  
                        {\cal H}_a & {\cal H}_b \\
                        {\cal H}_b^* & {\cal H}_a \\          
              \end{array} \right)
\label{matrix}
\end{eqnarray}

Where $\hat {\sigma}_x$ is a Pauli matrix.   The $3\times 3$
matrixes ${\cal H}_a$ and ${\cal H}_b$ are 
\begin{eqnarray}
{\cal H}_a &=& \left( \begin{array}{ccc}  
                        2 & 1+e^{i\left(k_x-k_y\right)} & 
                            -2+2e^{-ik_z}\\
                        1+e^{i\left(k_y-k_x\right)} & 2 & 
                             -2+2e^{-ik_z}\\
                        -2+2e^{ik_z} &  -2+2e^{ik_z}  & 8-8\cos k_z\\ 
            \end{array} \right)
\nonumber\\
{\cal H}_b &=& \left( \begin{array}{ccc}  
                        -2\cos k_x & -e^{-ik_y}-e^{ik_x} & 
                            2e^{ik_x}-2e^{i\left(k_x-k_z\right)}\\
                         -e^{ik_y}-e^{-ik_x}  & -2\cos k_y & 
                              2e^{ik_y}-2e^{i\left(k_y-k_z\right)} \\
                      2e^{-ik_x}-2e^{i\left(k_z-k_x\right)} &
                      2e^{-ik_y}-2e^{i\left(k_z-k_y\right)}  & 0\\ 
            \end{array} \right)
\end{eqnarray}

By solving (\ref{matrix}) for eigenvalues and eigenvectors we find 
the dispersion 
law of orbitons and phonons. In our approximation it is consistent to neglect 
$\hbar \omega$ relative to $\Delta$
in the denominators of (\ref{matrix}). Then we obtain two 
dispersive modes for orbitons and two for phonons, while 4 phonon 
modes remains degenerate with eigenvalues $E_{ph}^{3,4,5,6}=\hbar \omega$. 
The dispersive modes are:
\begin{eqnarray}
E_{\rm orb}^1&=&\frac{11}{12}\Delta+\frac{\hbar\omega}{12}
\left(6-4\cos k_z+\cos k_x+\cos k_y\right)
\nonumber\\
E_{\rm orb}^2&=&\frac{11}{12}\Delta+\frac{\hbar\omega}{12}
\left(6-4\cos k_z-\cos k_x-\cos k_y\right)
\nonumber\\
E_{\rm ph}^1&=&\hbar\omega-\frac{\hbar\omega}{12}
\left(6-4\cos k_z+\cos k_x+\cos k_y\right)
\nonumber\\
E_{\rm ph}^2&=&\hbar\omega-\frac{\hbar\omega}{12}
\left(6-4\cos k_z-\cos k_x-\cos k_y\right)
\label{dsp}
\end{eqnarray}

\begin{figure}
\centerline{
\psfig{figure=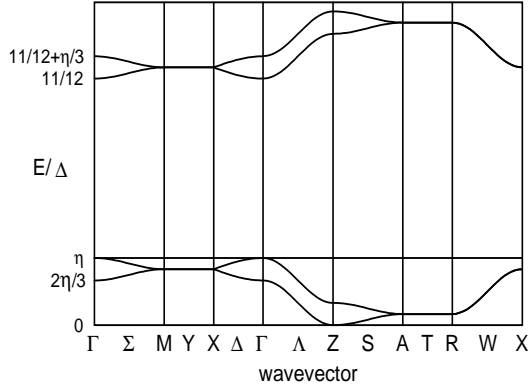,height=2.6in,width=2.6in,angle=0}}
\caption{Dispersion relation of orbitons (upper two curves) and phonons
of Jahn-Teller distorted LaMnO$_3$.  Energies are measured in units
of the JT gap $\Delta$.  Four phonon branches decouple from orbitons
and retain their bare energies $\hbar\omega$, independent of wavevector.
The parameter $\eta=\hbar\omega/\Delta$
has been set to 0.25 for display purposes.  A likely value of this parameter
is 0.06.}
\label{disp}
\end{figure}

The Jahn-Teller distortion $Q_2$ with a $(\pi,\pi,0)$ wavevector creates 
a simple tetragonal Brillouin zone, show in Fig. \ref{BZ}.
Fig. \ref{disp} shows the orbiton and phonon eigenvalues along 
principle symmetry lines.  An interesting
result is that the energy of one of 
the phonon branches goes to zero at wavevector  $\vec{k}=(0,0,\pi)$. 
The ground 
state energy of our Hamiltonian is degenerate in $(Q_2, Q_3)$ plane. 
It depends 
only on magnitude of the vector $\vec{Q}=(Q_2,Q_3)$, but not on its direction. 
We have chosen a $Q_2$-type distortion as is observed in experiment.
In our model, an infinitesimal distortion of the $Q_3$ type does not
cost any energy.  This Goldstone mode occurs necessarily at the $Z$ 
point of the Brillouin zone because only a $\vec{q}=(\pi,\pi,\pi)$
order parameter is independent of the direction of ordering in
$(Q_2,Q_3)$-space. 

\section{conclusion}

In this work we discuss the dispersion of the orbiton excitation caused by 
coupling to phonons. We solve a simple but reasonably realistic
model, which has previously been
used \cite{AP,Millis} to describe the insulating nature and 
magnetic ordering of lightly doped  LaMnO$_3$. The solution makes a 
single occupancy approximation, allowing only one electron per Mn site. 
We restrict consideration to creation of only a single orbiton defect.
We show that the electron-phonon interaction reduces the 
energy of the excitation by a 
factor of two, which is due to lattice relaxation. We also obtain a zero 
energy excitation or Goldstone mode which is due to a broken continuous
symmetry of the Hamiltonian.  We ignored two other sources of dispersion
of the orbiton -- the hopping terms ($t^2/U$), and direct quadrupolar
Coulomb coupling.  These effects would probably cause a similar
size dispersion to the one we considered, but would not have the major
effect of reducing the JT gap by a factor of two which occurs because of
coupling to phonons.

\acknowledgements
We thank A. J. Millis, J. P. Hill and S. Ishihara  for help.
This work was supported in part by NSF grant no. DMR-9725037.

\end{document}